\documentclass[twocolumn,prb]{revtex4}
\usepackage{amsfonts}
\usepackage[T1]{fontenc}
\usepackage{amsmath,amsbsy,amssymb,graphicx}
\usepackage{times}

\begin{document}

\title{Strong and weak second-order topological insulators with hexagonal
symmetry and $\mathbb{Z}_3$ index}
\author{Motohiko Ezawa}
\affiliation{Department of Applied Physics, University of Tokyo, Hongo 7-3-1, 113-8656,
Japan}

\begin{abstract}
We propose second-order topological insulators (SOTIs) whose lattice
structure has the hexagonal symmetry $C_{6}$ in three and two dimensions. We
start with a three-dimensional weak topological insulator constructed on the
stacked triangular lattice, which has only side topological surface states.
We then introduce an additional mass term which gaps out the side surface
states but preserves the hinge states. The resultant system is a
three-dimensional SOTI. The bulk topological quantum number is shown to be
the $\mathbb{Z}_{3}$ index protected by the inversion time-reversal symmetry 
$IT$ and the rotoinversion symmetry $C_{6}I$. We obtain three phases;
trivial, strong and weak SOTI phases. We argue the origin of these two types
of SOTIs. A hexagonal prism is a typical structure respecting these
symmetries, where six topological hinge states emerge at the side. The
building block is a hexagon in two dimensions, where topological corner
states emerge at the six corners in the SOTI phase. Strong and weak SOTIs
are obtained when the interlayer hopping interaction is strong and weak,
respectively. They are characterized by the emergence of hinge states
attached to or detached from the bulk bands.
\end{abstract}

\maketitle

Topological phase of matter remains to be a most active field of condensed
matter physics. Topological insulators (TIs) are well established, where the
emergence of topological boundary states is a manifestation of the bulk
topological number\cite{Hasan,Qi}. This is known as the bulk-boundary
correspondence. Recently, the notion of topological insulators is
generalized to second-order topological insulators (SOTIs)\cite%
{Fan,Science,APS,Peng,Lang,Song,Bena,Schin,FuRot,EzawaKagome,EzawaPhos,Gei,Kha}%
. A SOTI is such an insulator that has no topological surface states though
it has topological hinge states. They are one-dimensional (1D) edge states
emerging at hinges of a prism respecting the symmetry based upon which the
bulk topological quantum number is defined.

One powerful method to create a SOTI is to introduce a mass term to a strong
TI in such a way that it gaps out surface states but preserves hinge states%
\cite{Schin}. A topological hinge insulator was first constructed\cite{Schin}
by applying this method to the $C_{4}$ symmetric lattice model. However, in
this model\cite{Schin} a tetragonal prism of finite size has gapless surface
states at the top and the bottom of the prism in addition to four gapless
hinge states. This is because the symmetry indicator is characterized by $%
C_{4}T$ and $\bar{C}_{4}=C_{4}I$, where $T$ and $I$ are the time-reversal
symmetry generator and the inversion generator, respectively. These surface
states can be gapped out\cite{MagHOTI} by introducing the Zeeman term
violating $C_{4}T$ but preserving $\bar{C}_{4}$. The bulk topological
number, being characterized by the rotoinversion symmetry $\bar{C}_{4}$, is
given by the $\mathbb{Z}_{2}$ index.

\begin{figure}[t]
\centerline{\includegraphics[width=0.5\textwidth]{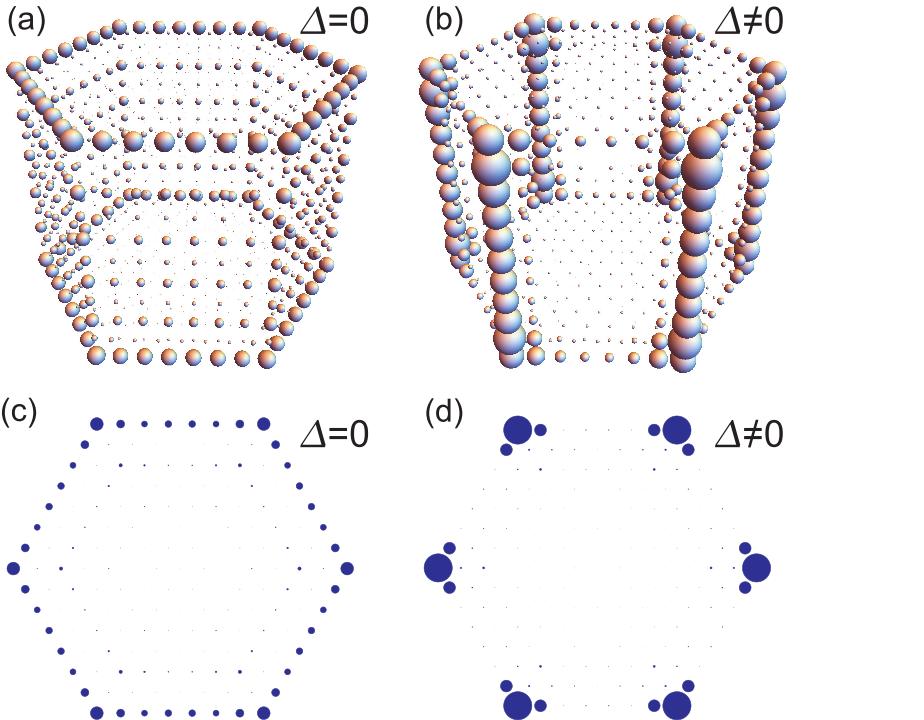}}
\caption{ (a)--(b) A hexagonal prism is a typical structure respecting the
rotoinversion symmetry $\bar{C}_{6}$. The real-space plot of the local
density of states $\protect\rho _{i}$ for a hexagonal prism in the case of
(a) a weak TI ($\Delta =0$) and (b) a topological hinge insulator ($\Delta
=0.7$). The amplitude is represented by the radius of spheres. The length of
one side of the hexagon is $L=8$, and the height of the prism is $H=11$.
(c)--(d) A hexagon is a typical structure respecting the rotoinversion
symmetry $\bar{C}_{6}$. The real-space plot of $\protect\rho _{i}$ for a
hexagon in the case of (c) a TI and (d) a topological corner insulator. We
have set $t=1$, $t_{z}=2$, $\protect\lambda =\protect\lambda _{z}=1$ and $%
m=3 $. }
\label{FigPrism}
\end{figure}

Very recently, a SOTI was experimentally materialized in Bismuth\cite%
{BisHOTI}, by employing topological quantum chemistry for material prediction%
\cite{TQC1,TQC2,TQC3,TQC4,TQC5,TQC6}. It has the $C_{6}$ symmetric lattice
structure, while the bulk topological quantum number, being characterized by 
$C_{3}$, $T$ and $I$, is given by the $\mathbb{Z}_{2}$ index. The
tight-binding model has been proposed, but it is an eight-band model and
rather too complicated.

\begin{figure*}[t]
\centerline{\includegraphics[width=1.00\textwidth]{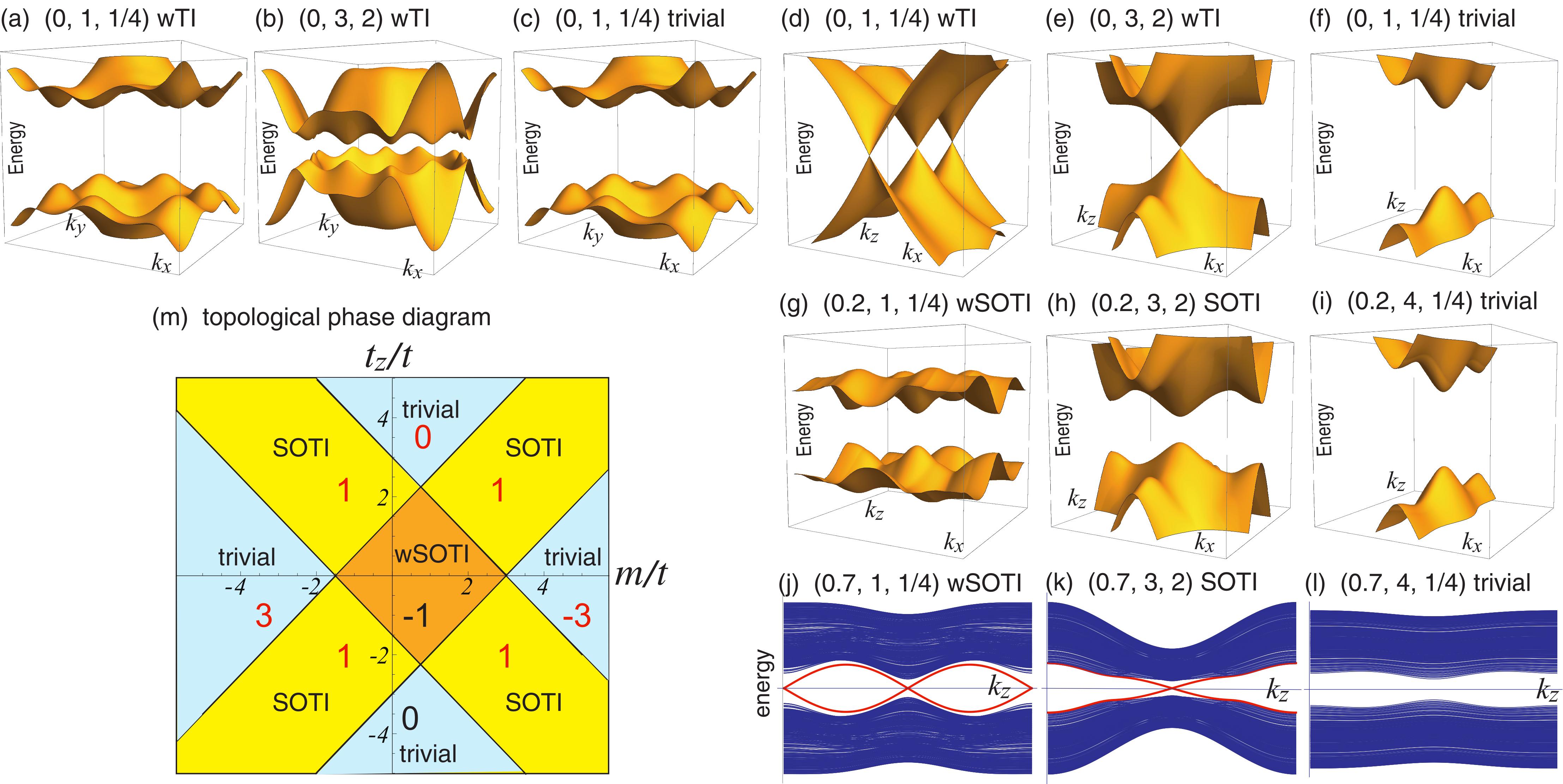}}
\caption{(a)--(c) Band structures of top surface states in a thin film for $%
\Delta =0$, where gaps are open. (d)--(f) Band structures of side surface
states in a thin film for $\Delta \not=0$. Two and one Dirac cones are
observed in (d)--(e), respectively, showing the system is in the weak TI
phase. A gap exists in (f), showing the system is in the trivial phase.
(g)--(i) The figures corresponding to (d)--(f) for $\Delta \not=0$, showing
the weak TI phase is transformed into the SOTI phase. (j)--(l) Band
structures of a hexagonal prism, where hinge states are shown in red. They
are remnants of th Dirac cones observed in (d) and (e). Parameters $(\Delta
,m,t_{z})$ are displayed in figures. The other parameters are $t=1=\protect%
\lambda =\protect\lambda _{z}$=1. (m) Topological phase diagram of the 3D
model. The horizontal axis is $m/t$, while the vertical axis is $t_{z}/t$.
Numbers in red represent the symmetry indicator $\protect\kappa _{6}$. The
topological quantum number $\protect\nu _{\text{3D}}$ is defined by $\protect%
\nu _{\text{3D}}=\text{mod}_{3}\protect\kappa _{6}$. Thus, there are two
SOTI phases shown by $\protect\kappa _{6}=\pm 1$, corresponding to (j) and
(k). It does not depend on $\protect\lambda ,\protect\lambda _{z}$ and $%
\Delta $. The horizontal axis of (j)--(l) is $k_{z}$ in the range of $(-%
\protect\pi ,\protect\pi )$.}
\label{FigSurface}
\end{figure*}

In this paper, we propose a simple four-band model possessing the $C_{6}$
symmetric lattice structure realizing a SOTI. We start with a weak TI
realized on the stacked triangular lattice, whose topological surface states
are present only at the side surfaces. Namely, it has no gapless surface
states at the top and the bottom when we consider a hexagonal prism of
finite size as in Fig.\ref{FigPrism}(a). Then, we gap out the side surface
states by introducing an additional mass term with parameter $\Delta $. As a
result, we obtain a SOTI, which has topological hinge states as in Fig.\ref%
{FigPrism}(b). The bulk topological quantum number is shown to be the $%
\mathbb{Z}_{3}$ index characterized by combinations of the rotoinversion
symmetry $\bar{C}_{6}=C_{6}I$ and the inversion time-reversal symmetry $IT$.
In accord with the $\mathbb{Z}_{3}$ index, there exist two different types
of SOTIs. We call them strong and weak SOTIs.

\textit{2D Hamiltonian: }We start with the 2D Hamiltonian\cite{EzawaNJP} 
\begin{equation}
H_{\text{2D}}^{0}=H_{t}\tau _{z}+H_{\text{SO}}\tau _{x}  \label{Hamil2D}
\end{equation}%
on the triangular lattice, where 
\begin{align}
H_{t}& =\sum_{n=1}^{3}\left[ m-t\sum \cos \left( \mathbf{d}_{n}\cdot \mathbf{%
k}\right) \right] , \\
H_{\text{SO}}& =\lambda \sum_{n=1}^{3}C_{3}^{n}\sigma _{x}C_{3}^{-n}\sin
\left( \mathbf{d}_{n}\cdot \mathbf{k}\right)
\end{align}%
in the momentum space. Here, $m$, $t$, $\lambda $ are real parameters, $%
\mathbf{k}=\left( k_{x},k_{y}\right) $, and $\mathbf{d}_{n}=|\mathbf{d}%
_{n}|[\cos (2\pi n/3),\sin (2\pi n/3)]$ is the pointing vector; $\mathbf{%
\sigma }=(\sigma _{x},\sigma _{y},\sigma _{z})$ and $\mathbf{\tau }=(\tau
_{x},\tau _{y},\tau _{z})$ represent the Pauli matrices for the spin and the
pseudospin corresponding to the orbital degrees of freedom, respectively; $%
C_{3}=\tau _{0}\exp \left[ -i\pi \sigma _{z}/3\right] $ is the generator of
the $\pi /3$ rotation. It has been shown\cite{EzawaNJP} that the Hamiltonian 
$H_{\text{2D}}^{0}$ describes a 2D TI. According to the bulk-boundary
correspondence there emerge gapless edge states for a hexagon as in Fig.\ref%
{FigPrism}(c).

We propose to consider the Hamiltonian%
\begin{equation}
H_{\text{2D}}=H_{\text{2D}}^{0}+H_{\Delta }\tau _{0}
\end{equation}%
with%
\begin{equation}
H_{\Delta }=\Delta \sum_{n=1}^{3}C_{3}^{n}\sigma _{y}C_{3}^{-n}\cos \left( 
\mathbf{d}_{n}\cdot \mathbf{k}\right) ],
\end{equation}%
where $\Delta $ is a real parameter. This term is a hexagonal generalization
of the term proposed for the tetragonal symmetric system\cite{Schin}. Its
role is to gap out edge states except for the corners, leading to a 2D
topological corner insulator as in Fig.\ref{FigPrism}(d), about which we
discuss at the end of this paper.

\textit{3D Hamiltonian: }We may stack triangular lattices to generate a 3D
lattice. The relevant Hamiltonian is given by 
\begin{equation}
H_{\text{3D}}^{0}=H_{\text{2D}}^{0}+H_{z}\quad \text{or}\quad H_{\text{3D}%
}=H_{\text{2D}}+H_{z}
\end{equation}
depending whether $\Delta =0$ or $\Delta \neq 0$, with 
\begin{equation}
H_{z}=-t_{z}\tau _{z}\cos k_{z}+\lambda _{z}\tau _{x}\sin k_{z},
\end{equation}%
which is a well-known term describing interlayer hopping with real
parameters $t_{z}$ and the spin-orbit interaction $\lambda _{z}$.

We consider a hexagonal prism of finite size subject to $H_{\text{3D}}^{0}$.
Provided both $t_{z}$ and $\lambda _{z}$ are small enough, since the prism
is simply obtained by stacking hexagons having the density of state as in
Fig.\ref{FigPrism}(c), it has naturally topological surface states at the
six sides but none at the top and bottom as in Fig.\ref{FigPrism}(a). This
remains true even if the parameters $t_{z}$ and $\lambda _{z}$ are not small
based on explicit calculations. See Fig.\ref{FigSurface}(a)--(c) with
respect to the absence of gapless surface states at the top and the bottom.
Such an insulator is called a weak TI. The problem is how to gap out the
side surface states preserving the hinge states. This is the main issue of
the present work. Our result demonstrates that the $H_{\Delta }\tau _{0}$
term transforms the weak TI into a topological hinge insulator as in Fig.\ref%
{FigPrism}(b).

\textit{Topological phase diagram:} We analyze the Hamiltonian $H_{\text{3D}%
} $. The band structure is obtained by diagonalizing it. The essential point
is that the bulk topological quantum number is defined by the band structure
at the six high-symmetry points with respect to the six-fold rotoinversion:
See (\ref{TC-C4I}) with (\ref{Kappa4}). They are $\Gamma =\left(
0,0,0\right) $, $K=\left( 4\pi /3,0,0\right) $, $K^{\prime }=(-4\pi /3,0,0)$%
, $A=\left( 0,0,\pi \right) $, $H=\left( 4\pi /3,0,\pi \right) $, $H^{\prime
}=(-4\pi /3,0,\pi )$. Consequently, to determine the topological phase
boundaries, it is enough to solve the zero-energy condition ($E=0$) at these
six high-symmetry points.

The energies at these points are analytically obtained as 
\begin{align}
E\left( \Gamma \right) & =\pm \left( 3t+t_{z}-m\right) , \\
E\left( K\right) & =E\left( K^{\prime }\right) =\pm \left(
3t/2-t_{z}+m\right) , \\
E\left( A\right) & =\pm \left( 3t-t_{z}-m\right) , \\
E\left( H\right) & =E\left( H^{\prime }\right) =\pm \left(
3t/2+t_{z}+m\right) .
\end{align}%
The topological phase boundary is given by $t_{z}=\pm 3t+m$ and $t_{z}=\pm
3t/2-m$, which are shown in Fig.\ref{FigSurface}(m). They are independent of
the values of $\lambda ,\lambda _{z},\Delta $.

\textit{Symmetries:} In order to identify the bulk topological quantum
number, it is necessary to study the symmetry of the Hamiltonian $H_{\text{3D%
}}$. We note that the Hamiltonian $H_{\text{3D}}^{0}$ has both the
time-reversal symmetry $TH_{\text{3D}}^{0}\left( \mathbf{k}\right) T^{-1}=H_{%
\text{3D}}^{0}\left( -\mathbf{k}\right) $ and the inversion symmetry $IH_{%
\text{3D}}^{0}\left( \mathbf{k}\right) I^{-1}=H_{\text{3D}}^{0}\left( -%
\mathbf{k}\right) $, where $T=-i\tau _{0}\sigma _{y}K$ generates the time
reversal symmetry (TRS) with the complex conjugation $K$, while $I=\tau
_{z}\sigma _{0}$ is the inversion symmetry generator. In addition, there is
the six-fold rotational symmetry $C_{6}$, 
\begin{equation}
C_{6}H_{\text{3D}}^{0}\left( k_{x},k_{y},k_{z}\right) C_{6}^{-1}=H_{\text{3D}%
}^{0}\left( k_{x}^{\prime },k_{y}^{\prime },k_{z}\right) ,
\end{equation}%
where $C_{6}=\tau _{0}\exp \left[ -i\pi \sigma _{z}/6\right] $ is the
generator of the $\pi /6$ rotation, and%
\begin{equation}
k_{x}^{\prime }=k_{x}/2+\sqrt{3}k_{y}/2,\quad k_{y}^{\prime }=-\sqrt{3}%
k_{x}/2+k_{y}/2.
\end{equation}%
The term $H_{\Delta }$ breaks both the TRS and the inversion symmetry but
preserves the combined symmetry $IT$ and the rotoinversion symmetry $\bar{C}%
_{6}=C_{6}I$, which is similar to the case of the tetragonal system\cite%
{Schin}.

\textit{Symmetry indicator:} The symmetry indicator is already known for the
tetragonal system possessing the $\bar{C}_{4}$ and $IT$ symmetries\cite%
{Schin,k4}. By making its hexagonal generalization, we define the symmetry
indicator $\kappa _{6}$ protected by the $\bar{C}_{6}$ and $IT$ symmetries
by the formula 
\begin{equation}
\kappa _{6}=\frac{1}{2\sqrt{3}}\sum_{k}\sum_{\alpha }e^{\frac{i\alpha \pi }{6%
}}n_{K}^{\alpha },  \label{Kappa4}
\end{equation}%
where $k$ runs over the symmetry invariant points associated with the $\bar{C%
}_{6}$, $\Gamma $, $K$, $K^{\prime }$, $A$, $H$ and $H^{\prime }$; $%
n_{K}^{\alpha }$ is the number of the occupied bands with the eigenvalue $e^{%
\frac{i\alpha \pi }{6}}$ of the symmetry operator $\bar{C}_{6}$, $\bar{C}%
_{6}|\psi \rangle =e^{\frac{i\alpha \pi }{6}}|\psi \rangle $. The symmetry
indicator is shown to be quantized and real. First, $\alpha $ is quantized
to be $\alpha =1$, $3$, $5$, $7$, $9$, $11$, because of the relation $\left( 
\bar{C}_{6}\right) ^{6}=-1$. Second, $\kappa _{6}$ is real, since the band
structure is always two-fold degenerated in the presence of the $IT$
symmetry. The symmetry eigenvalues form a conjugate pair\cite{Schin} for
these bands due to the commutation relation $[\bar{C}_{6},IT]=0$ and the
fact that the $IT$ symmetry is anti-unitary. Third, $\kappa _{6}$ is a
constant within one topological phase since it can change its value only
when a phase boundary is crossed by changing system parameters. Consequently
it is a candidate of the topological quantum number. We explicitly evaluate $%
\kappa _{6}$ using the formula (\ref{Kappa4}), which is shown in the phase
diagram Fig.\ref{FigSurface}(m).

\begin{figure}[t]
\centerline{\includegraphics[width=0.5\textwidth]{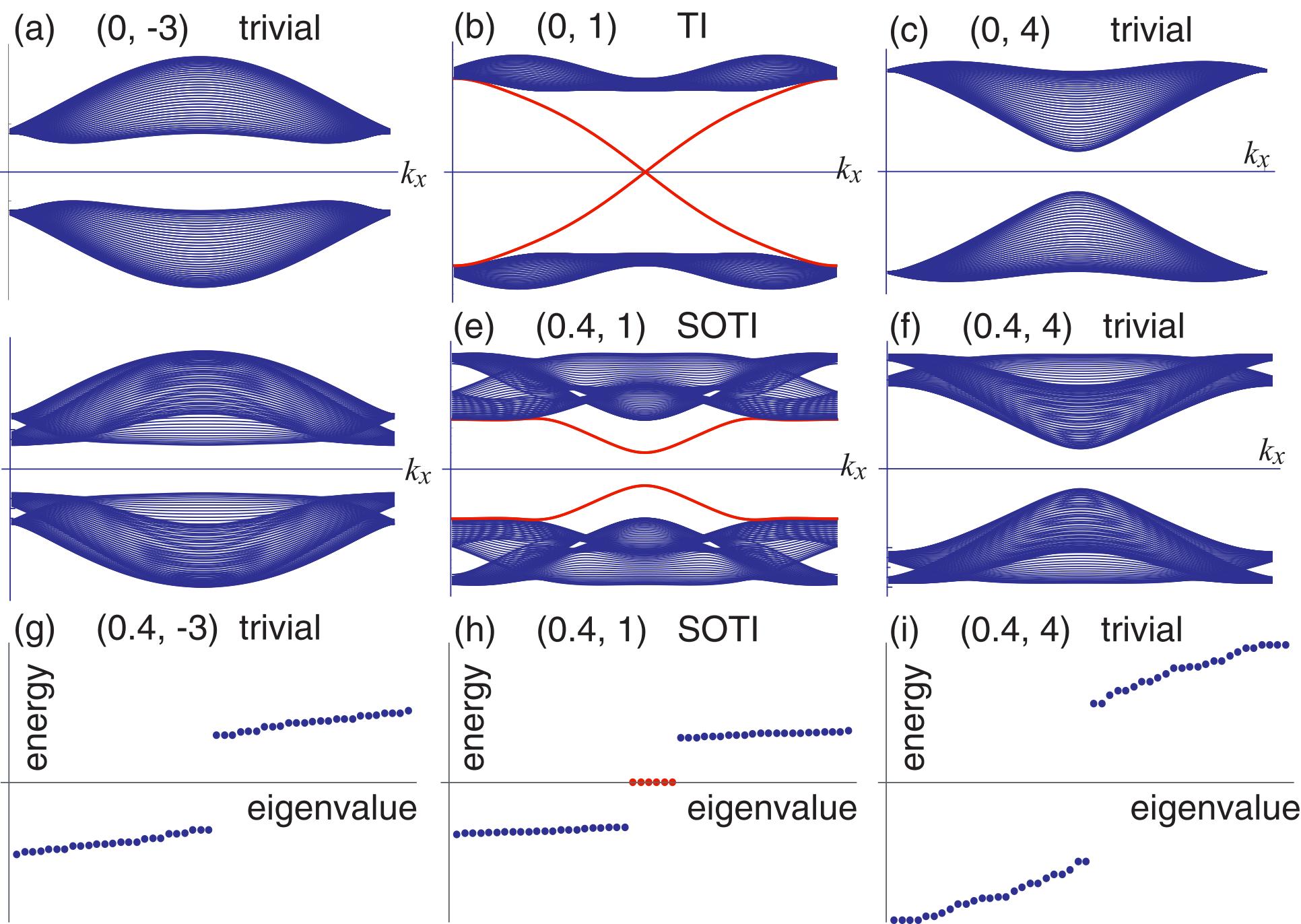}}
\caption{Band structures of the 2D model (a)--(c) with $\Delta =0$ and
(d)--(f) with $\Delta \not=0$ in nanoribbon geometry for typical points in
the phase diagram. (g)--(i) Eigenenergy of the 2D model in hexagonal
nanodisk geometry for typical points in the phase diagram. (h) Six
degenerate zero-energy states emerge in the SOTI phase. Parameters $(\Delta
,m)$ are displayed in figures. The horizontal axis of (a)--(f) is $k_{x}$ in
the range of $(-\protect\pi ,\protect\pi )$.}
\label{FigHexa}
\end{figure}

\textit{Surface states:} We first examine the surface states by analyzing
the band structure of a thin film. (i) When $\Delta =0$, gapless modes are
found in the side surfaces [Fig.\ref{FigSurface}(d)--(e)] in the TI phase,
but not in the up and bottom surfaces [Fig.\ref{FigSurface}(a)--(b)], which
shows that the system is a weak TI. (ii) When $\Delta \not=0$, these surface
states are gapped [Fig.\ref{FigSurface}(g)--(h)]. These features are
consistent with the local density of states for the hexagonal prism as shown
in Fig.\ref{FigPrism}(a)--(b). Hence, the system is naively a trivial
insulator according to the bulk-boundary correspondence. However, as we now
see, hinge states appears.

\textit{Hinge states and $\mathbb{Z}_{3}$ index protected by $\bar{C}_{6}$
and $IT$:} We next investigate the hinge states by analyzing the band
structure of a hexagonal prism [Fig.\ref{FigSurface}(j)--(l)] at various
points in the phase diagram [Fig.\ref{FigSurface}(m)]. According to the band
structure of a hexagonal prism, hinge states emerge in the two phases
indexed by $\kappa _{6}=\pm 1$ as in Fig.\ref{FigSurface}(j)--(k), while no
hinge states are present in the phase indexed by $\kappa _{6}=0,\pm 3$ as in
Fig.\ref{FigSurface}(l). They are identified with SOTI and trivial phases.
The two SOTI phases indexed by $\kappa _{6}=\pm 1$ are distinguished by the
band structure of hinge states. Namely, hinge states are detached from
(attached to) the bulk band for $\kappa _{6}=-1$ ($\kappa _{6}=1$): See Fig.%
\ref{FigSurface}(j) and (k). Consequently, the bulk topological index is
given by the $\mathbb{Z}_{3}$ index defined by 
\begin{equation}
\nu _{\text{3D}}=\text{mod}_{3}\kappa _{6},  \label{TC-C4I}
\end{equation}%
which is a generalization of the $\mathbb{Z}_{2}$ index $\nu _{0}$ into the
hexagonal symmetric system in the absence of the TRS and the inversion
symmetry.

It is intriguing that we have two different types of hinge states. We may
understand their origin as follows. Recall that the hexagonal prism is
described by the Hamiltonian $H_{\text{3D}}=H_{\text{2D}}+H_{z}$. The
building block of a hexagonal prism is a hexagon described by $H_{\text{2D}}$%
. As we soon discuss, it has six detached corner states protected
topologically, as shown in Fig.\ref{FigPrism}(d). Hence, when we construct a
prism by stacking hexagons, we would obtain a perfect flat band for the
hinge states in the vanishing limit of interlayer hopping ($H_{z}\rightarrow
0$). Since the flat band is deformed solely by the interlayer hopping
interaction, we expect that such hinge states are determined by the
Hamiltonian $H_{z}$. Indeed, the band structure of the detached hinge states
is obtained analytically by diagonalizing the Hamiltonian $H_{z}$,\ and
given by%
\begin{equation}
E=\pm \frac{1}{\sqrt{2}}\sqrt{t_{z}^{2}+\lambda _{z}^{2}+\left(
t_{z}^{2}-\lambda _{z}^{2}\right) \cos 2k_{z}}.
\end{equation}%
This solution well reproduces the detached hinge states in Fig.\ref%
{FigSurface}(j). This is the origin of hinge states in the SOTI phase with $%
\kappa _{6}=-1$, which we call a weak SOTI phase. On the other hand, when
the interlayer hopping interaction is strong, a mixing occurs between $H_{%
\text{2D}}$ and $H_{z}$, making the corner states a part of the bulk bands.
The resultant hinge states form the SOTI phase with $\kappa _{6}=1$, which
we call a strong SOTI phase.

\textit{2D SOTI:} We study a hexagonal SOTI model in two dimensions. The
Hamiltonian is $H_{\text{2D}}$ given by (\ref{Hamil2D}). The symmetry
analysis is almost the same as in the 3D case just by neglecting the $z$
coordinate. The properties of the 2D hexagonal SOTI are summarized as
follows.

First, the topological phase diagram is obtained by setting $t_{z}=0$ in Fig.%
\ref{FigSurface}(m). Namely, there emerge a SOTI phase for $-3/2<m/t<1$, and
trivial phases for $m/t<-3/2$ and $m/t>3$. Second, the $\kappa _{6}$ index
is obtained as $\kappa _{6}=3/2$ for $m/t<-3/2$, $\kappa _{6}=-1/2$ for $%
-3/2<m/t<1$ and $\kappa _{6}=-3/2$ for $m/t>3$. Hence, we define the bulk
topological quantum number as 
\begin{equation}
\nu _{\text{2D}}=\text{mod}_{3}(2\kappa _{6}).  \label{TC-C4Ib}
\end{equation}%
Third, we show the band structure of a nanoribbon in Fig.\ref{FigHexa}. When 
$\Delta =0$, there are helical edge states in Fig.\ref{FigHexa}(b). We have
previously shown\cite{EzawaNJP} that the system is a TI. They are gapped for 
$\Delta \neq 0$ as in Fig.\ref{FigHexa}(e), indicating that the system would
be topologically trivial. Actually, it is not trivial but it is in the SOTI
phase. Indeed, when we evaluate the eigenvalues of a hexagonal nanodisk,
six-fold degenerate zero-energy corner states emerge for the SOTI phase as
in Fig.\ref{FigHexa}(h), showing the system is a topological corner
insulator as in Fig.\ref{FigPrism}(d).

In this work we have presented a simple model for a hexagonal topological
hinge insulator. Although the hinge structure at the sides looks the same as
that of Bismuth, there are some difference between them. Indeed, the
topological quantum number is the $\mathbb{Z}_{3}$ index in the present
model but it is the $\mathbb{Z}_{2}$ index in the Bismuth model. The origin
of the difference is traced back to the fact that a hexagonal prism is
constructed by stacking hexagons which are weak TIs. It yields two types of
SOTIs depending on the interlayer hopping interaction whether it is strong
or weak.

The author is very much grateful to B. A. Bernevig, T. Neupert and N.
Nagaosa for helpful discussions on the subject. This work is supported by
the Grants-in-Aid for Scientific Research from MEXT KAKENHI (Grant
Nos.JP17K05490 and JP15H05854). This work is also supported by CREST, JST
(JPMJCR16F1).


\begin{thebibliography}{99}
\bibitem{Hasan} M. Z. Hasan and C. L. Kane, Rev. Mod. Phys. \textbf{82},
3045 (2010).

\bibitem{Qi} X.-L. Qi and S.-C. Zhang, Rev. Mod. Phys. \textbf{83}, 1057
(2011).

\bibitem{Fan} F. Zhang, C.L. Kane and E.J. Mele, Phys. Rev. Lett. \textbf{110%
}, 046404 (2013).

\bibitem{Science} W. A. Benalcazar, B. A. Bernevig, and T. L. Hughes,
10.1126/science.aah6442.

\bibitem{APS} F. Schindler, A. Cook, M. G. Vergniory, and T. Neupert, in APS
March Meeting (2017).

\bibitem{Peng} Y. Peng, Y. Bao, and F. von Oppen, Phys. Rev. B \textbf{95},
235143 (2017).

\bibitem{Lang} J. Langbehn, Y. Peng, L. Trifunovic, F. von Oppen, and P. W.
Brouwer, Phys. Rev. Lett. \textbf{119}, 246401 (2017).

\bibitem{Song} Z. Song, Z. Fang, and C. Fang, Phys. Rev. Lett. \textbf{119},
246402 (2017).

\bibitem{Bena} W. A. Benalcazar, B. A. Bernevig, and T. L. Hughes, Phys.
Rev. B \textbf{96}, 245115 (2017).

\bibitem{Schin} F. Schindler, A. M. Cook, M. G. Vergniory, Z. Wang, S. S. P.
Parkin, B. A. Bernevig, and T. Neupert, cond-mat/arXiv:1708.03636 (2017). .

\bibitem{FuRot} C. Fang, L. Fu, arXiv:1709.01929

\bibitem{EzawaKagome} M. Ezawa, Phys. Rev. Lett. 120, 026801 (2018)

\bibitem{EzawaPhos} M. Ezawa, cond-mat/arXiv:1801.00437.

\bibitem{Gei} M. Geier, L. Trifunovic, M. Hoskam, and P. W. Brouwer,
cond-mat/arXiv:1801.10053

\bibitem{Kha} E. Khalaf, cond-mat/arXiv:1801.10050

\bibitem{MagHOTI} M. Ezawa, cond-mat/arXiv:1802.03571.

\bibitem{BisHOTI} F. Schindler, Z. Wang, M. G. Vergniory, A. M. Cook, A.
Murani, S. Sengupta, A. Y. Kasumov, R. Deblock, S. Jeon, I. Drozdov, H.
Bouchiat, S. Gueon, A. Yazdani, B. A. Bernevig, and T. Neupert,
cond-mat/arXiv:1802.02585.

\bibitem{TQC1} B. Bradlyn, L. Elcoro, J. Cano, M. G. Vergniory, Z. Wang, C.
Felser, M. I. Aroyo, and B. A. Bernevig, Nature 547, 298 (2017).

\bibitem{TQC2} M. G. Vergniory, L. Elcoro, Z. Wang, J. Cano, C. Felser, M.
I.Aroyo, B. A. Bernevig, and B. Bradlyn, Phys. Rev. E 96, 023310 (2017).

\bibitem{TQC3} L. Elcoro, B. Bradlyn, Z. Wang, M. G. Vergniory, J. Cano, C.
Felser, B. A. Bernevig, D. Orobengoa, G. Flor, and M. I. Aroyo, Journal of
Applied Crystallography 50 (2017).

\bibitem{TQC4} J. Cano, B. Bradlyn, Z. Wang, L. Elcoro, M. G. Vergniory, C.
Felser, M. I. Aroyo, and B. A. Bernevig, cond-mat/arXiv:1709.01935

\bibitem{TQC5} B. Bradlyn, L. Elcoro, M. G. Vergniory, J. Cano, Z. Wang, C.
Felser, M. I. Aroyo, and B. A. Bernevig, ArXiv e-prints (2017),
cond-mat/arXiv:1709.01937

\bibitem{TQC6} J. Cano, B. Bradlyn, Z. Wang, L. Elcoro, M. G. Vergniory, C.
Felser, M. I. Aroyo, and B. A. Bernevig, cond-mat/arXiv:1711.11045

\bibitem{EzawaNJP} M. Ezawa, New J. Phys. 16, 065015 (2014).

\bibitem{k4} E. Khalaf, H. C. Po, A. Vishwanath and H. Watanabe,
cond-mat/arXiv:1711.11589.
\end{thebibliography}
\end{document}